\documentclass[12pt,a4paper,twocolumn]{narms2}
\usepackage{amsmath}
\usepackage{graphicx}
\newcommand{\be}{\begin{equation}}
\newcommand{\ee}{\end{equation}}
\newcommand{\bi}{\begin{itemize}}
\newcommand{\ei}{\end{itemize}}

\usepackage{subfigure} % was: subfignarms.sty
\usepackage{psfig}
\usepackage{timesmt}   % use times typeface
\usepackage{chikako}   % <---------------------------------- MOD

\makeatletter
\renewcommand{\section}{\@startsection%
{section}%
{1}%
{0mm}%
{- \baselineskip}%
{0.15\baselineskip}%
{\normalfont\normalsize}}%

\renewcommand{\subsection}{\@startsection
{subsection}%
{2}%
{0mm}%
{-\baselineskip}%
{0.15\baselineskip}%
{\normalfont\normalsize}}%
\makeatother

\begin{document}
%\maketitle

\title{The nature of quasistatic deformation in granular materials}
\author{\large {J.-N. Roux}\\
{\em Laboratoire des Mat\'eriaux et des Structures du G\'enie Civil, Institut Navier, 
Champs-sur-Marne, France}\\
}
\date{}% No date.
\abstract{ABSTRACT: Strain in granular materials in quasistatic conditions under varying stress originate 
in (I) contact deformation and (II) rearrangements of the contact network. 
Depending on sample history and applied load, either mechanism might dominate. 
One may thus define rheological regimes I and II accordingly.
Their properties are presented and illustrated here with discrete numerical simulation results on sphere packings. 
Understanding the microscopic physical origin of strain enables one to clarify such issues 
as the existence of macroscopic elasticity, the approach to stress-strain relations in the large system 
limit and the sensitivity to noise. 
}
%%%%%%%%%%%%%%%%%%%%%%%%%%%%%%%%%%%%%%%%%%%%%%%%%%%%%%%%%%%%%%%%%%%
\maketitle
\frenchspacing  
\section{INTRODUCTION}
Macroscopic strain in solidlike granular materials has two obvious
physical origins: first, grains deform near their contacts, where
stresses concentrate (so that one models intergranular interaction
with a point force); then, grain packs rearrange as contact networks,
between two different equilibrium states break, and then repair in a
different stable configuration. We refer here respectively to the two
different kinds of strains as type I and II. The purpose of the
present communication is to delineate the regimes, denoted as I and II
accordingly, within which one mechanism or the other dominates, in a
simple model material (an assembly of spheres), from discrete
numerical simulations. Macroscopic mechanical properties are shown
to differ, as well as microscopic variables.   

The very small strain elastic
response of granular materials belongs to regime I: what is measured
then is the macroscopic stiffness of a spring network, each
intergranular contact behaving like an elastic element. Such a spring
network model is usually adopted on studying vibration modes and
elastic moduli \shortcite{Somfai,Ivana-ici}.
However, elastic-frictional contact networks also comprise
plastic elements (sliders), and deform
irreversibly under quasistatically applied stress increments. As long
as they still support the applied load, strain amplitudes scale as the
inverse of the stiffness constants of the springs. Such a scaling will
be used here as a signature of regime I, which extends, beyond the
quasi-elastic domain, throughout the
stress or strain interval corresponding to the elastoplastic response
of a given contact network. If grains are modeled as perfectly rigid,
strains in regime I reduce to zero.

Regime II will in general correspond to larger strains, for which contact networks keep rearranging.
Strain amplitudes are then related to the distances (gaps) between
neighbouring grains that do not touch. Contact stiffnesses are then
expected to have little influence on macroscopic deformations. Such
situations are sometimes studied by simulation methods that deal
with rigid grains, such as Contact Dynamics \shortcite{RR04}. As the
network continuously fails and repairs, larger dynamical effects and
larger spatial fluctuations of strain are expected,
since failing materials usually exhibit larger heterogeneities.

The simulations reported below explore the conditions of occurrence of
regimes I and II, and give a quantitative meaning to the statements made in this
introduction. After basic features of numerical computations are introduced
in Sec.~\ref{sec:basic}, results on the constitutive law and regimes I and II are given in Sec.~\ref{sec:3D}.   
Sec.~\ref{sec:concl} is a brief conclusion.
\section{NUMERICAL MODEL\label{sec:basic}}
Triaxial compression tests of monosized assemblies of $N$ ($N=4000$
for most results here).
spheres of diameter $a$ were simulated by molecular dynamics (MD, or DEM).
In those computer experiments, one starts
with an isotropically assembled initial state with pressure $P$, and
then, keeping the axes of coordinates as principal stress directions,
increases slowly the largest principal stress, $\sigma_1$, while the others are held fixed, equal to $P$. 
One denotes as $q$ the stress deviator, $q=\sigma_1-p$. Like in most numerical studies, we chose here to impose 
a constant strain rate $\dot \epsilon _1$ and to measure $\sigma_1$ as a function of $\epsilon_1$, termed
``axial strain'' and subsequantly denoted as  $\epsilon_a$. Soil mechanics conventions are adopted: compressive
stresses and shrinking strains are positive.
We focus on the quasistatic mechanical behaviour expressed by dependences $q(\epsilon_a)$,
$\epsilon_v(\epsilon_a)$ as $\epsilon_a$ increases, in dense systems, before the deviator peak is reached.
$\epsilon_v$ is the \emph{volumetric strain} (relative volume
decrease). Dimensional analysis leads to the definition of the 
inertia parameter $I = \dot \epsilon_a\sqrt{m/aP}$ as a measure of the
departure from equilibrium, the quasistatic limit being $I\to 0$.

Motivated by possible comparisons to laboratory
experiments with glass bead packings, simulations are carried
out with Hertz-Mindlin contacts, with the elastic properties of glass (Young modulus
$E=70$~GPa, Poisson coefficient $\nu=0.3$), and a friction coefficient $\mu =0.3$ -- 
additional details and references are provided in \shortcite{Ivana-ici}. Normal viscous forces
are also implemented: the damping parameter in any contact is chosen as a fixed fraction $\zeta$ of its critical value, 
defined for the contacting pair with its instantaneous (i.e., dependent on current normal force or $h$) stiffness constant $dF_N/dh$.
A suitable dimensionless parameter characterizing the importance of elastic deflections $h$ in contacts is 
$\kappa = (E/P)^{2/3}$ (such that $h/a \sim 1/\kappa$).
\section{SIMULATION RESULTS\label{sec:3D}}
\subsection{{\em Sample preparation}}
\begin{figure}[!htb]
\centering
%\fbox{
\includegraphics[width=7.5cm]{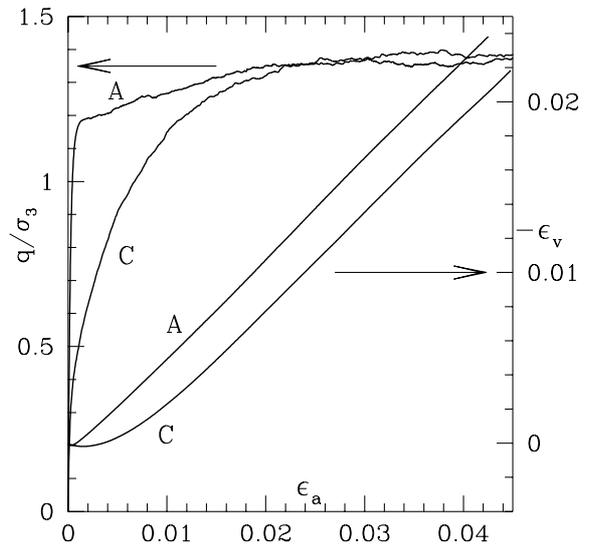}
%}
\caption{ 
%{\small{\bf
$q(\epsilon_a)$ (left scale)  and $\epsilon_v(\epsilon_a)$ (right scale) curves for A and C 
states under $P=100$~kPa. Averages over 5 samples of 4000 spherical grains.
%}}
\label{fig:triaxcomp}}
\end{figure}

In order to obtain dense samples, we first simulated sets
of sphere packings prepared by isotropic compression
of frictionless granular gases. This results in configurations
hereafter denoted as A. A-type configurations have a high coordination number (approaching 6 at low
pressure if inactive grains are discarded). They therefore present a
large force indeterminacy. We observed (fig.~\ref{fig:triaxcomp}) that the raise of deviator
$q$ with axial strain in such samples is much faster than in usual
experimental results, for which $\epsilon_a$ is usually of order $1\%$ to $5\%$ at the deviator peak.
Likewise, the onset of dilatancy after the initial contractant strain interval is unusually fast in A samples.
 This motivated the use of a different 
preparation procedure, which, although idealized, aims to imitate the effects of vibrations 
in the assembling of a dense, dry packing of beads. In this method (called C in the sequel),
A samples are first dilated (multiplying coordinates by $1.005$), then mixed, as by thermal agitation,
until each grain has had 50 collisions on average, and finally compressed in the presence of friction to
a relativy low pressure, $P=10kPa$. Higher $P$ values are obtained on further compressing. 
Fig.~\ref{fig:triaxcomp} compares
the behaviour of initial states A and C, in triaxial compression
with $P=100$~kPa ($\kappa \simeq 6000$).
\shortciteN{Ivana-ici}
report in these proceedings on the large difference in coordination number between A
and C states, where it is much smaller ($\sim 4.7$), while densities
are very close. Usual 
experimental curves, which do not exhibit $q$ maxima or dilatancy before $\epsilon_a \sim 0.01$,
are better modelled with C samples. Those experiments are made with,
e.g., dry grains assembled in the laboratory. One cannot exclude,
however, that samples left to age and anneal for a long time gradually
evolve towards better coordinated configurations resembling A
ones. One may also assemble the grains in the
presence of a lubricant, thereby strongly reducing friction in the  
initial stage \shortcite{Xiaoping-ici}. A-type samples can thus be
viewed as ideal models for preparation procedures suppressing
friction, while C ones are more appropriate models for
laboratory specimens made by pouring, vibrating or tapping. A similar
conclusion was reached by \shortciteN{Xiaoping-ici} in a study of sound propagation velocities.
\subsection{{\em Reproducibility, quasistatic limit}}
\begin{figure}[!htb]
\centering
\includegraphics[height=7.5cm,angle=270]{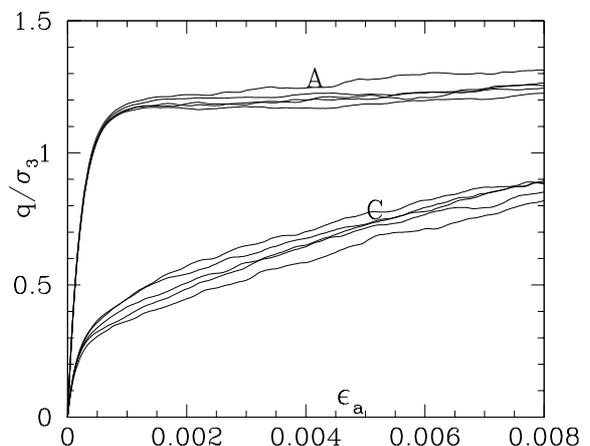}
\caption{Detail of small strain part of $q(\epsilon_a)$ curves for 5
  different
samples of each type, A (top curves) and C (bottom ones) with $N=4000$
beads.
\label{fig:triaxcomp2}
}
\end{figure}

\begin{figure}[!htb]
\centering
\includegraphics[width=7.5cm]{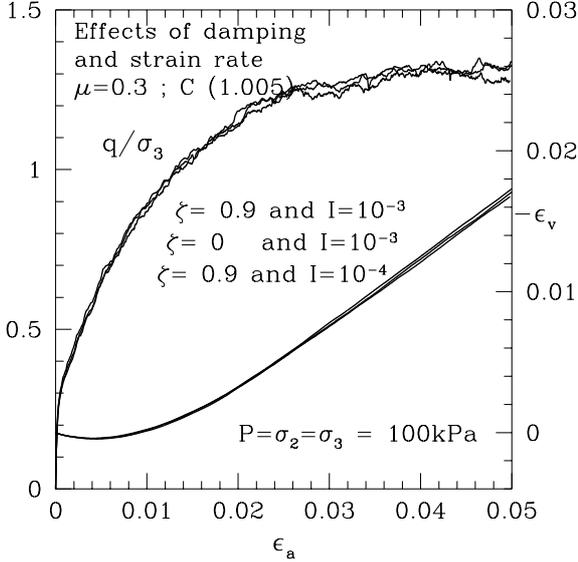}
\caption{Effect of dynamical parameters: $q(\epsilon_a)$ and $\epsilon_v(\epsilon_a)$ curves for the different values of
$\zeta$ and $I$ indicated coincide, showing the innocuousness of dynamical parameter choice.
\label{fig:visc}
}
\end{figure}

Stress-strain curves as displayed on fig.~\ref{fig:triaxcomp} should express a macroscopic, 
quasistatic constitutive law. Sample to sample fluctuations should regress in the large system limit, 
and the results should be independent on
dynamical parameters such as inertia, viscous dissipation, and strain rate, summarized in dimensionless parameters $\xi$ and $I$.
Fig.~\ref{fig:triaxcomp2} is indicative of sample to sample fluctuations with 4000 beads. One may notice the 
very good reproducibility of the curve between A samples in the
initial fastly growing part. 
We checked that differences between samples
decreased for increasing $N$. As to the influence of dynamical
parameters, 
fig.~\ref{fig:visc} shows that the quasistatic limit is correctly
approached for $I\le 10^{-3}$, a quite satisfactory result, given that usual laboratory tests with $\dot \epsilon_a \sim 10^{-5}$
correspond to $I\le 10^{-8}$.
In previous 2D simulations with disks \shortcite{RC02}, sample to
sample fluctuations were shown to regress as $N^{-1/2}$.
\subsection{{\em Influence of contact stiffness}}
\begin{figure}[!htb]
\centering
\includegraphics[width=7.5cm]{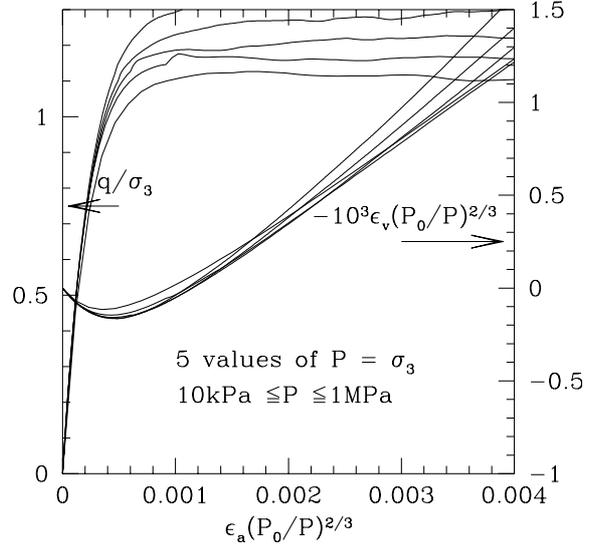}
\caption{ 
%{\small{\bf
$q(\epsilon_a)/P$ and $\epsilon_v(\epsilon_a)$ curves for one A sample and different
$P$ values. Strains on scale $(P/P_0)^{2/3}\propto \kappa^{-1}$, $P_0=100$~kPa.
%}}
\label{fig:dessPP2}}
\end{figure}

\begin{figure}[!htb]
\centering
\includegraphics[width=7.5cm]{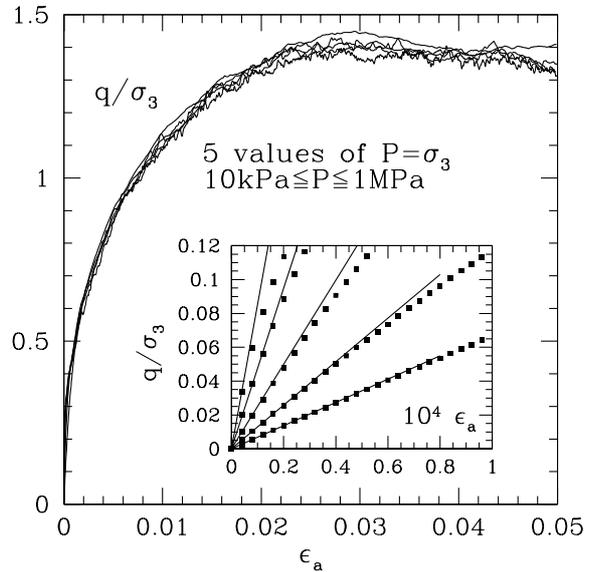}
\caption{ 
%{\small{\bf
$q(\epsilon_a)/p$ for same $p$ values as on fig.~\ref{fig:dessPP2}, in one $C$ sample. Inset: detail of same
curves, blown-up $\epsilon$ scale, straight lines corresponding to Young moduli in isotropic state.
%}}
\label{fig:dessP1}}
\end{figure}

The small strain (say $\epsilon_a\le 5.10^{-4}$) interval for A samples, with its fast $q$ increase, is in fact in regime I.
This is readily checked on changing the confining pressure. Fig.~\ref{fig:dessPP2} shows 
the curves for triaxial compressions at different $P$ values
(separated by a factor $\sqrt{10}$) from $10$~kPa to $1$~MPa,
with a rescaling of the strains by the stiffness parameter $\kappa$,
in one A sample. 
Their coincidence for $q/P\le 0.8P$ evidences a 
wide deviator range in regime I. 
For larger strains, curves separate on this scale, and tend to collapse together if $q/P$, $\epsilon_v$ are simply plotted versus
$\epsilon_a$. The strain
dependence on stress ratio is independent from contact stiffness. This different sensitivity to pressure
is characteristic of regime II.
Fig~\ref{fig:dessP1} shows that it applies to C samples almost throughout the investigated
range, down to small deviators (a behaviour closer to usual
experimental results than type A configurations). At the origin (close to the
initial isotropic state, see inset on fig.~\ref{fig:dessP1}), the
tangent to the curve is given by the elastic (Young) modulus of the
granular material, $E_m$, which scales as $\kappa$, but curves 
quickly depart from this behaviour (below $q=0.1P$). 
\subsection{{\em Load reversal}}
\begin{figure}[!htb]
\centering
\includegraphics[width=8cm]{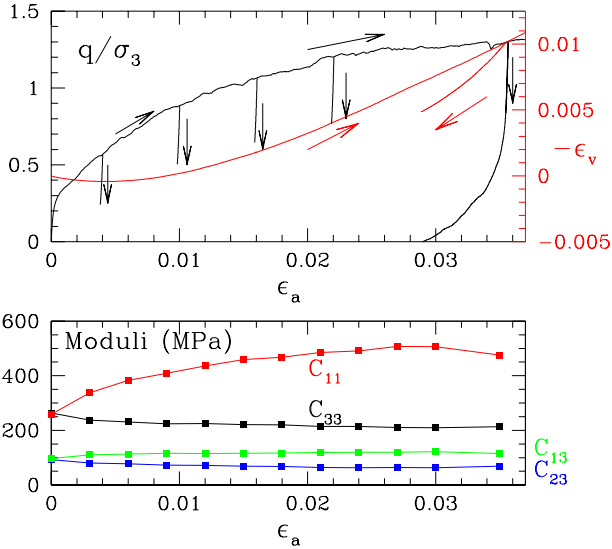}
\caption{ 
%{\small{\bf
Top plot: effects of load reversals at different points on curves (C
sample).
Initial slopes of unloading curves
correspond to elastic moduli. 
Bottom: evolution with $\epsilon_a$ of some elastic moduli,
probing induced anisotropy.
%}}
\label{fig:dessJM}}
\end{figure}

If (fig.~\ref{fig:dessJM}) one reverses the
direction of load increments, the stress-strain curves exhibit notable
intervals
within which the deviator stress decreases very fast, which
results in large irreversible (plastic) strains. 
It can be checked that the initial slope of those descending curves are equal to the Young modulus $E_m$ 
of the material, and that subsequent strains scale as $1/E_m$, like the initial $q$ increase in A samples.  
Therefore, some significant deviator stress intervals (of order $0.2P$ or larger) are found in regime I
on reversing deviator stress or axial strain variations.  

\begin{figure}[!htb]
\centering
\includegraphics[height=8cm,angle=270]{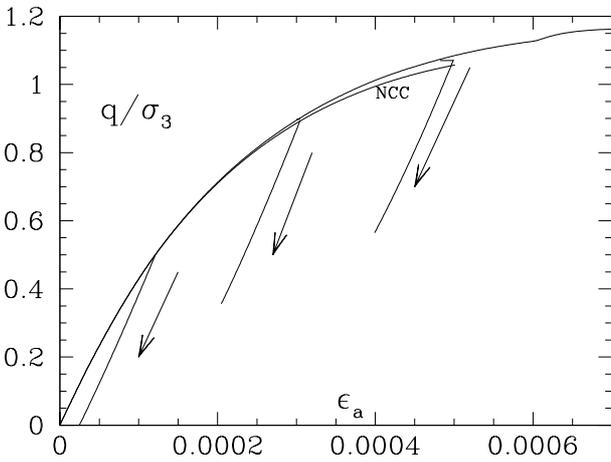}
\caption{Very small strain part of $q(\epsilon_a)$ curve in one A sample, showing beginning of unloading curves (arrows).
Curve marked  NCC was obtained on calculating the evolution of the same sample without any contact creation. \label{fig:revers}
}
\end{figure}

Fig.~\ref{fig:revers} shows that the small strain response of A samples, within regime I, close to the initial state,
is already irreversible. Type I
strains are not elastic.
\subsection{{\em Calculations with a fixed contact list}}
Within regime I, the mechanical properties of the material can be
successfully
predicted on studying the response of one given set of contacts.
Those might slide or open, but the very few new contacts that are
created can be neglected.
To check this in simulations, 
one may restrict at each time step the search for interacting grains
to the list of initially contacting pairs.
Fig.~\ref{fig:revers} compares such a procedure to the complete
calculation. 
The curve marked ``NCC'' for \emph{no
  contact creation} is
indistinguishable from the other one for $q\ge 0.8$. In two dimensions, \shortciteN{RC02}
could implement a purely static method (elastoplastic computation on a
given contact network), apt to calculate the quasistatic evolution of the sample
under varying applied stresses throughout the initial regime I stage
of 2D assemblies of disks analogous to A samples. The limit between
regimes I and II was studied with some accuracy \shortcite{GaelERLPC},
and shown to approach a finite value in the rigid limit ($\kappa \to
+\infty$), and in the limit of large systems. This value does not
appear to depend on details of 
contact elasticity, such as tangential to normal stiffness ratio \shortcite{GaelERLPC}.
\subsection{{\em Microscopic aspects}}
The existence of wide stress intervals within regime I is associated
with strongly hyperstatic contact networks (large force
indeterminacy). Initially, A samples have large coordination numbers,
(close to 6) \shortcite{Ivana-ici}, and friction is not mobilized (zero
tangential forces). Consequently, the set of contact forces that
resolve the load and satisfy Coulomb inequalities is large, and the
initial forces are far from its boundaries. At coordination 6 this set spans an affine
space of dimension $3N^*$ if $N^*$ is the number of force-carrying particles.
In regime II, regarding the Coulomb condition in sliding contacts as a
constraint on force values in the count of force indeterminacy, this
dimension decreases to a fraction of order $10\%$ of the number of
degrees of freedom. Upon reversing the load variation, sliding
contacts tend to disappear, leading to a larger force indeterminacy
and a notable type I interval.
The small variation of coordination number in the
pressure range of Fig.~\ref{fig:dessP1}
\shortcite{Xiaoping-ici} witnesses the smallness of geometrical
changes, hence the collapse of curves with type II strains.

Larger strain heterogeneities and sensitivity to perturbations are
other characteristic features of regime II \shortcite{RC03}. 
\section{CONCLUSION\label{sec:concl}}
Numerical studies thus reveal that the two regimes, in which the origins of
strain differ, exhibit contrasting properties. On attempting to predict
a macroscopic mechanical response from packing geometry and contact
laws, the information about which kind of strain should dominate is
crucial. Regime I corresponds in usual testing conditions 
to highly coordinated systems (with many contacts), or to changes in the direction
of load increments (hence a loss in friction mobilization). Investigating the nature of strains might open 
interesting perspectives to study the effects of cyclic loadings or random perturbations.
\bibliographystyle{chikako}     
%\bibliography{../granu}

\end{document}